\documentclass[11pt]{article}

\usepackage[english]{babel}
\usepackage[latin1]{inputenc}
\usepackage{amsmath, latexsym}
\usepackage[psamsfonts]{amsfonts}
\usepackage{amssymb}
\usepackage[authoryear]{natbib}

\newif\ifpdf
\ifx\pdfoutput\undefined
   \pdffalse
\else
   \pdfoutput=1
   \pdftrue
\fi

\ifpdf
    \usepackage{color}
    \definecolor{myred}{rgb}{0.5,0,0}
    \definecolor{myblue}{rgb}{0,0,0.75}
    \definecolor{mygreen}{rgb}{0,0.5,0}
   \usepackage[pdftex,
               pdftitle={Estimating Probabilities of Default for Low Default Portfolios},
               pdfsubject={Low default portfolios},
               pdfkeywords={},
               pdfauthor={Katja Pluto, Dirk Tasche},
               bookmarks=false,
               breaklinks=true,
               colorlinks=true,
               citecolor=myred,
               linkcolor=myblue,
               urlcolor=mygreen]{hyperref}
\else
    \newcommand{\href}[1]{}
\fi

\newtheorem{theorem}{Theorem}[section]

\newtheorem{proposition}[theorem]{Proposition}

\numberwithin{equation}{section}

\newlength{\captionwidth}
\begin{document}

\title{Estimating Probabilities of Default for Low Default Portfolios%
}

\author{%
Katja Pluto and Dirk Tasche\thanks{Deutsche Bundesbank, Postfach
10 06 02, 60006 Frankfurt am
Main, Germany\newline E-mail: katja.pluto@gmx.de, dirk.tasche@gmx.net}\ %
\thanks{The opinions expressed in this note are those of the
authors and do not necessarily reflect views of the Deutsche Bundesbank.} }

\date{April 4, 2005}
\maketitle

\begin{abstract}
For credit risk management purposes in general, and for allocation
of regulatory capital by banks in particular (Basel II), numerical
assessments of the credit-worthiness of borrowers are
indispensable. These assessments are expressed in terms of
probabilities of default (PD) that should incorporate a certain
degree of conservatism in order to reflect the prudential risk
management style banks are required to apply. In case of credit
portfolios that did not at all suffer defaults, or very few
defaults only over years, the resulting naive zero or close to
zero estimates would clearly not involve such a sufficient
conservatism. As an attempt to overcome this issue, we suggest the
\emph{most prudent estimation} principle. This means to estimate
the PDs by upper confidence bounds while guaranteeing at the same
time a PD ordering that respects the differences in credit quality
indicated by the rating grades. The methodology is most easily
applied under an assumption of independent default events but can
be adapted to the case of correlated defaults.
\end{abstract}


\section{Introduction}
A core input to modern credit risk modeling and managing
techniques are probabilities of default (PD) per borrower. As
such, the accuracy of the PD estimations determines the quality of
the results of credit risk models.

One of the obstacles connected with PD estimations can be the low
number of defaults, especially in the better rating grades. Good
rating grades might experience many years without any defaults.
And even if some defaults occur in a given year, the observed
default rates might exhibit a high degree of volatility, due to
the relatively low number of borrowers in that grade. But even
entire portfolios with low or no defaults are not uncommon in
reality. Examples include portfolios with an overall good quality
of borrowers (e.g.\ sovereign or bank portfolios) as well as
high-volume-low-number portfolios (e.g.\ specialized lending).

Usual bank practices for deriving PD values for such exposures
often focus on qualitative mapping mechanisms to bank-wide master
scales or external ratings. These practices, while widespread in
the industry, do not entirely satisfy the desire for a statistical
foundation of the assumed PD values. One may ``believe'' that the
PDs per rating grade appear correct, as well as believe that the
ordinal ranking and the relative spread between the PDs of two
grades is right, but information about the absolute PD figures is
lacking. Lastly, it could be questioned whether these rather
qualitative methods of PD calibration fulfill the minimum
requirements set out in \citet{BC04a}.

The issue has, amongst others, recently been raised in
\citet{BBA04}. In that paper, applications of causal default
models and of exogenous distribution assumptions on the PDs across
the grades have been proposed. In a recent paper,
\citet{SchuermannHanson04} present a methodology to estimate PDs
by means of migration matrices \citep[``duration method'', cf.\
also][]{JafrySchuermann04}. This way, non-zero PDs for
high-quality rating grades can be estimated more precisely by
counting the borrower migrations through the lower grades to
eventual default and using Markov chain properties.

This paper focuses on a different issue of PD estimations in low
default portfolios. We present a methodology to estimate PDs for
portfolios without any defaults, or a very low number of defaults
in the overall portfolio. The proposal by
\citeauthor{SchuermannHanson04} does not provide a solution for
such cases, because the duration method requires a certain number
of defaults in at least some (usually the low-quality) rating
grades.

For estimating PDs, we use all available quantitative information
of the rating system and its grades. Moreover, we assume that the
ordinal borrower ranking is correct. We do not use any additional
assumptions or information. Every additional piece of input would
be more on the assumption side, as the low default property of
these portfolios does not provide us with more reliable quantitative
information.

Our methodology delivers confidence intervals for the PDs of each
rating grade. The PD range can be adjusted by the choice of an
appropriate confidence level. Moreover, by the \emph{most prudent
estimation} principle our methodology yields monotone PD
estimates. We look both at the cases of uncorrelated and
correlated default events, in the latter case under assumptions
consistent with the Basel risk weight model.

Moreover, we extend the \emph{most prudent estimation} by two
application variants: First we scale our results to overall
portfolio central tendencies. Second, we apply our methodology to
multi-period data and extend our model by time dependencies of the
Basel systematic factor. Both variants should help to align our
principle to realistic data sets and to a range of assumptions
that can be set according to the specific issues in question when
applying our methodology.

The paper is structured as follows: The two main concepts underlying
the methodology --
estimating PDs as upper confidence bounds and guaranteeing their
monotony by the most prudent estimation principle --
are introduced by two examples that assume
independence of the default events. The first example deals with a
portfolio without any observed defaults. For the second example,
we modify the first example by assuming that a few defaults have
been observed.
In a further section, we show how the methodology can be modified
in order to take into account non-zero correlation of default
events. This is followed by two sections discussing potential
extensions of our methodology, in particular the scaling to the
overall portfolio central tendency and an extension of our model
to the multi-period case. The last two sections are devoted to
discussions of the potential scope of application and of open
questions. We conclude with a summary of our proposal. In Appendix
\ref{sec:appendix}, we provide information on the numerics that is
needed to implement the estimation approach we suggest. Appendix
\ref{sec:appendix2} provides additional numerical results to
Section \ref{sec:scaling}.


\section{Example: No Defaults, Assumption of Independence}
\label{sec:2}
The obligors are distributed to rating grades $A$, $B$, and $C$, with
frequencies $n_A$, $n_B$, and $n_C$. The grade with the highest
credit-worthiness is denoted by $A$, the grade with the lowest credit-worthiness
is denoted by $C$. Neither in $A$ nor in $B$ nor in $C$ any defaults occurred
during the last observation period.

We assume that the -- still to be estimated -- PDs $p_A$ of grade $A$, $p_B$
of grade $B$, and $p_C$ of grade $C$ reflect the decreasing credit-worthiness
of the grades, in the sense of the following inequality:
\begin{equation}
  \label{eq:ineq}
p_A \ \le \ p_B \ \le \ p_C.
\end{equation}
The inequality implies that we assume the ordinal borrower ranking
to be correct. According to \eqref{eq:ineq}, the PD $p_A$ of grade
$A$ cannot be greater than the PD $p_C$ of grade $C$. As a
consequence, the \emph{most prudent
  estimate} of the value of $p_A$ is obtained under the assumption that the
probabilities $p_A$ and $p_C$ are equal. Then, from \eqref{eq:ineq} even
follows $p_A = p_B = p_C$. Assuming this relation, we now proceed in
determining a confidence region for $p_A$ at confidence level $\gamma$. This
confidence region\footnote{%
For any value of $p_A$ \emph{not} belonging to this region, the hypothesis
that the true PD takes on this value would have to be rejected at a type I
error level of $1-\gamma$.
} can be described as the set of all admissible values of $p_A$ with the
property that the probability of not observing any default during the
observation period is not less than $1-\gamma$ (for instance for $\gamma =
90\%$).

If we have got $p_A = p_B = p_C$, then the three rating grades
$A$, $B$, and $C$ do not differ in their respective riskiness.
Hence we have to deal with a homogeneous sample of size $n_A + n_B
+ n_C$ without any default during the observation period. Assuming
unconditional independence of the default events, the probability
of observing no defaults turns out to be $(1-p_A)^{n_A + n_B +
n_C}$. As a consequence, we have to solve the inequality
\begin{subequations}
  \begin{equation}\label{eq:2a}
    1-\gamma  \le (1-p_A)^{n_A + n_B + n_C}
\end{equation}
for $p_A$ in order to obtain the confidence region at level $\gamma$ for $p_A$
as the set of all the values of $p_A$ such that
\begin{equation}
p_A  \le
1-(1-\gamma)^{1/(n_A+n_B+n_C)}.\label{eq:2b}
  \end{equation}
\end{subequations}
If we choose for the sake of illustration
\begin{equation}
  \label{eq:3}
  n_A = 100, \quad n_B = 400, \quad n_C = 300,
\end{equation}
Table \ref{tab:1} exhibits some values of confidence levels
$\gamma$ with the corresponding maximum values (upper confidence
bounds) $\hat{p}_A$ of $p_A$ such that \eqref{eq:2a} is still
satisfied.
  \begin{table}[htbp]
    \begin{center}
\parbox{\captionwidth}{\caption{ \label{tab:1}\emph{Upper confidence bound
      $\hat{p}_A$ of $p_A$ as a
function of the
  confidence level $\gamma$. No defaults observed, frequencies of obligors in
  grades given by \eqref{eq:3}.}}}
\begin{tabular}{c||c|c|c|c|c|c}
$\gamma$ & 50\% & 75\% & 90\% & 95\% & 99\% & 99.9\% \\ \hline
$\hat{p}_A$ & 0.09\% & 0.17\% & 0.29\% & 0.37\% & 0.57\% & 0.86\%
\end{tabular}
    \end{center}
  \end{table}

According to Table~\ref{tab:1}, there is a strong dependence of
the upper confidence bound $\hat{p}_A$ on the confidence level
$\gamma$. Intuitively, values of $\gamma$ smaller than  $95\%$
seem more appropriate for estimating the PD by $\hat{p}_A$.

By inequality \eqref{eq:ineq}, the PD $p_B$ of grade $B$ cannot be
greater than the PD $p_C$ of grade $C$ either. Consequently, the
\emph{most prudent estimate} of $p_B$ is obtained by assuming $p_B
= p_C$. Assuming additional equality with the PD $p_A$ of the best
grade $A$ would violate the \emph{most prudent estimation}
principle, because $p_A$ is a \emph{lower bound} of $p_B$. If we
have got $p_B = p_C$, then $B$ and $C$ do not differ in their
respective riskiness and may be considered a homogeneous sample of
size $n_B + n_C$. Therefore, the confidence region at level
$\gamma$ for $p_B$ is obtained from the inequality
\begin{subequations}
  \begin{equation}
    \label{eq:4a}
1-\gamma  \le (1-p_B)^{n_B + n_C}.
  \end{equation}
\eqref{eq:4a} implies that the confidence region for $p_B$ consists of all the
values of $p_B$ that satisfy
\begin{equation}
  \label{eq:4b}
p_B  \le 1-(1-\gamma)^{1/(n_B+n_C)}.
\end{equation}
\end{subequations}
If we take up again the example described by \eqref{eq:3}, Table
\ref{tab:2} exhibits some values of confidence levels $\gamma$
with the corresponding maximum values (upper confidence bounds)
$\hat{p}_B$ of $p_B$ such that \eqref{eq:4a} is still fulfilled.
  \begin{table}[htbp]
    \begin{center}
\parbox{\captionwidth}{\caption{ \label{tab:2}\emph{Upper confidence bound
      $\hat{p}_B$ of $p_B$ as a
function of the
  confidence level $\gamma$. No defaults observed, frequencies of obligors in
  grades given by \eqref{eq:3}.}}}
\begin{tabular}{c||c|c|c|c|c|c}
$\gamma$ & 50\% & 75\% & 90\% & 95\% & 99\% & 99.9\% \\ \hline
$\hat{p}_B$ & 0.10\% & 0.20\% & 0.33\% & 0.43\% & 0.66\% & 0.98\%
\end{tabular}
    \end{center}
  \end{table}

For determining the confidence region at level $\gamma$ for $p_C$
we only make use of the observations in grade $C$ because by
\eqref{eq:ineq} there is no obvious upper bound for $p_C$. Hence
the confidence region at level $\gamma$ for $p_C$ consists of
those values of $p_C$ that satisfy the inequality
\begin{subequations}
  \begin{equation}
    \label{eq:5a}
1-\gamma  \le (1-p_C)^{n_C}.
  \end{equation}
Equivalently, the confidence region for $p_C$ can be described by
\begin{equation}
  \label{eq:5b}
p_C  \le 1-(1-\gamma)^{1/n_C}.
\end{equation}
\end{subequations}
Coming back to our example \eqref{eq:3}, Table \ref{tab:3} lists
some values of confidence levels $\gamma$ with the corresponding
maximum values (upper confidence bounds) $\hat{p}_C$ of $p_C$ such
that \eqref{eq:5a} is still fulfilled.
  \begin{table}[htbp]
    \begin{center}
\parbox{\captionwidth}{\caption{ \label{tab:3}\emph{Upper confidence bound
      $\hat{p}_C$ of $p_C$ as a
function of the
  confidence level $\gamma$. No defaults observed, frequencies of obligors in
  grades given by \eqref{eq:3}.}}}
\begin{tabular}{c||c|c|c|c|c|c}
$\gamma$ & 50\% & 75\% & 90\% & 95\% & 99\% & 99.9\% \\ \hline
$\hat{p}_C$ &  0.23\% & 0.46\% & 0.76\% & 0.99\% & 1.52\% & 2.28\%
\end{tabular}
    \end{center}
  \end{table}

Comparison of Tables~\ref{tab:1}, \ref{tab:2}, and \ref{tab:3}
shows that -- besides the confidence level $\gamma$ -- the
applicable sample size is a main driver of the upper confidence
bound. The smaller the sample size that can be made use of, the
greater will be the upper confidence bound. This is not an
undesirable  effect because intuitively the credit-worthiness ought
to be the better, the greater is the number of obligors in a
portfolio without any default observation.

As the results presented so far seem plausible, we suggest to use
upper confidence bounds as described by \eqref{eq:2b},
\eqref{eq:4b}, and \eqref{eq:5b} as estimates for the PDs in
portfolios without observed defaults. The case of three rating
grades we have considered in this section can readily be
generalized to an arbitrary number of grades. We do not present
the details here.

However, the larger the number of obligors in the entire portfolio
is, the more often some defaults will occur in some grades at
least, even if the general quality of the portfolio is very high.
This case is not covered by \eqref{eq:2b}, \eqref{eq:4b}, and
\eqref{eq:5b}. In the following section, we will show -- still
keeping the assumption of independence of the default events --
how the \emph{most prudent
  estimation} methodology can be adapted to the case of a non-zero but still
low number of defaults.

\section{Example: Few Defaults, Assumption of Independence}
\label{sec:3}
We consider again the portfolio from Section~\ref{sec:2} with the frequencies
$n_A$, $n_B$, and $n_C$. In contrast to Section~\ref{sec:2}, this time we
assume that during the last period no default was observed in grade $A$, two
defaults were observed in grade $B$, and one default was observed in grade $C$.

As in Section~\ref{sec:2}, we determine a \emph{most prudent
confidence region} for the PD $p_A$ of $A$. Again, we do so by
assuming that the PDs of the three grades are equal. This allows
us to treat the entire portfolio as a homogeneous sample of size
$n_A + n_B + n_C$. Then the probability of observing not more than
three defaults is given by the expression
\begin{equation}\label{eq:p_A}
\sum_{i=0}^3
\left(\begin{smallmatrix}
  n_A+n_B+n_C\\ i
\end{smallmatrix}\right)
\,p_A^i\,(1-p_A)^{n_A+n_B+n_C-i}.
\end{equation}
\eqref{eq:p_A} follows from the fact that the number of defaults
in the portfolio is binomially distributed as long as the default
events are independent. As a consequence of \eqref{eq:p_A}, the
confidence region\footnote{We calculate the simple and intuitive
\emph{exact Clopper-Pearson interval}. For an overview of this
approach, as well as potential alternatives, see \citet{Brown01}.}
at level $\gamma$ for $p_A$ is given as the set of all the values
of $p_A$ that satisfy the inequality
\begin{equation}
  \label{eq:3defs}
  1-\gamma \ \le \ \sum_{i=0}^3
\left(\begin{smallmatrix}
  n_A+n_B+n_C\\ i
\end{smallmatrix}\right)
\,p_A^i\,(1-p_A)^{n_A+n_B+n_C-i}.
\end{equation}
The tail distribution of a binomial distribution can be expressed
in terms of an appropriate beta distribution function. Thus,
inequality \eqref{eq:3defs} can be solved
analytically\footnote{Alternatively, solving directly
\eqref{eq:3defs} for $p_A$ by means of numerical tools is not too
difficult either (see Appendix~\ref{sec:appendix},
Proposition~\ref{pr:1} for additional information).} for $p_A$.
For details, see Appendix~\ref{sec:appendix}. If we assume again
that the obligors' numbers per grade are as in \eqref{eq:3},
Table~\ref{tab:4} shows maximum solutions $\hat{p}_A$ of
\eqref{eq:3defs} for different confidence levels $\gamma$.
  \begin{table}[htbp]
    \begin{center}
\parbox{\captionwidth}{\caption{ \label{tab:4}\emph{Upper confidence bound
      $\hat{p}_A$ of $p_A$ as a
function of the
  confidence level $\gamma$. No default observed in grade $A$, two defaults
      observed in grade $B$, one default observed in grade $C$,
frequencies of obligors in
  grades given by \eqref{eq:3}.}}}
\begin{tabular}{c||c|c|c|c|c|c}
$\gamma$ & 50\% & 75\% & 90\% & 95\% & 99\% & 99.9\% \\ \hline
$\hat{p}_A$ & 0.46\% & 0.65\% & 0.83\% & 0.97\% & 1.25\% & 1.62\%
\end{tabular}
    \end{center}
  \end{table}

Although in grade $A$ no defaults have been observed, the three
defaults that occurred during the observation period enter the
calculation. They effect the upper confidence bounds, which are
much higher than those in Table~\ref{tab:1}. This is a consequence
of the precautionary assumption $p_A = p_B = p_C$. However, if we
alternatively considered grade $A$ alone (by reevaluating
\eqref{eq:5b} with $n_A = 100$ instead of $n_C$), we would obtain
an upper confidence bound $1.38\%$ at level $\gamma =
75\%$. This value is still much higher than the one that has been
calculated under the precautionary assumption $p_A = p_B = p_C$ --
a consequence of the low frequency of obligors in grade $A$ in
this example. Nevertheless, we see that the methodology described
by \eqref{eq:3defs} yields fairly reasonable results.

In order to determine the confidence region at level $\gamma$ for
$p_B$, as in Section ~\ref{sec:2}, we assume that $p_B$ takes its
greatest possible value according to \eqref{eq:ineq}, i.e.\ that
we have $p_B = p_C$. In this situation, we have got a homogeneous
portfolio with $n_B + n_C$ obligors, PD $p_B$, and three observed
defaults. In complete analogy to \eqref{eq:p_A}, the probability
of observing no more than three defaults in one period then can be
written as
\begin{equation}
  \label{eq:p_B}
\sum_{i=0}^3
\left(\begin{smallmatrix}
  n_B+n_C\\ i
\end{smallmatrix}\right)
\,p_B^i\,(1-p_B)^{n_B+n_C-i}.
\end{equation}
Hence, the confidence region at level $\gamma$ for $p_B$ turns out to be the
set of all the admissible values of $p_B$ which satisfy the inequality
\begin{equation}
  \label{eq:3defsB}
  1-\gamma \ \le \ \sum_{i=0}^3
\left(\begin{smallmatrix}
  n_B+n_C\\ i
\end{smallmatrix}\right)
\,p_B^i\,(1-p_B)^{n_B+n_C-i}.
\end{equation}
By analytically or numerically solving \eqref{eq:3defsB} for $p_B$
-- with frequencies of obligors in the grades as in \eqref{eq:3}
-- we obtain Table~\ref{tab:5} with some maximum solutions
$\hat{p}_B$ of \eqref{eq:3defsB} for different confidence levels
$\gamma$.
  \begin{table}[htbp]
    \begin{center}
\parbox{\captionwidth}{\caption{ \label{tab:5}\emph{Upper confidence bound
      $\hat{p}_B$ of $p_B$ as a
function of the
  confidence level $\gamma$. No default observed in grade $A$, two defaults
      observed in grade $B$, one default observed in grade $C$,
frequencies of obligors in
  grades given by \eqref{eq:3}.}}}
\begin{tabular}{c||c|c|c|c|c|c}
$\gamma$ & 50\% & 75\% & 90\% & 95\% & 99\% & 99.9\% \\ \hline
$\hat{p}_B$ & 0.52\% & 0.73\% & 0.95\% & 1.10\% & 1.43\% & 1.85\%
\end{tabular}
    \end{center}
  \end{table}

From the given numbers of defaults in the different grades it becomes clear
that a stand-alone treatment of grade $B$ would yield still much higher values\footnote{%
At level $99.9\%$, e.g., $2.78\%$ would  be the value of the upper
confidence bound. } for the upper confidence bounds. The upper
confidence bound $0.52\%$ of the confidence region at level $50\%$
is almost identical with the naive frequency based PD estimate
$2/400 = 0.5\%$ that could alternatively have been calculated for
grade $B$ in this example.

For determining the confidence region at level $\gamma$ for the PD
$p_C$, by the same rationale as in Section~\ref{sec:2} the grade
$C$ must be considered a stand-alone portfolio. According to the
assumption made in the beginning of this section, one default
occurred among the $n_C$ obligors in $C$. Hence we see that the
confidence region for $p_C$ is the set of all admissible values of
$p_C$ that satisfy the inequality
\begin{equation}
  \label{eq:3defsC}
  1-\gamma \ \le \ \sum_{i=0}^1
\left(\begin{smallmatrix}
  n_C\\ i
\end{smallmatrix}\right)
\,p_C^i\,(1-p_C)^{n_C-i}\ = \ (1-p_C)^{n_C} + n_C\,p_C\,(1-p_C)^{n_C-1}.
\end{equation}
For obligor frequencies as assumed in example
\eqref{eq:3}, Table~\ref{tab:6} exhibits some maximum solutions\footnote{%
If we had assumed that two defaults occurred in grade $B$ but no
default was observed in grade $C$, then we would have obtained
smaller upper bounds for $p_C$ than for $p_B$. As this is not a
desirable effect, a possible -- conservative -- work-around could
be to increment the number of defaults in grade $C$ up to the
point where $p_C$ would take on a greater value than $p_B$.
Nevertheless, in this case one would have to make sure that the
applied rating system yields indeed a correct ranking of the obligors.}
$\hat{p}_C$ of \eqref{eq:3defsC} for different
confidence levels $\gamma$.
\begin{table}[htbp]
    \begin{center}
\parbox{\captionwidth}{\caption{ \label{tab:6}\emph{Upper confidence bound
      $\hat{p}_C$ of $p_C$ as a
function of the
  confidence level $\gamma$. No default observed in grade $A$, two defaults
      observed in grade $B$, one default observed in grade $C$,
frequencies of obligors in
  grades given by \eqref{eq:3}.}}}
\begin{tabular}{c||c|c|c|c|c|c}
$\gamma$ & 50\% & 75\% & 90\% & 95\% & 99\% & 99.9\% \\ \hline
$\hat{p}_C$ & 0.56\% & 0.90\% & 1.29\% & 1.57\% & 2.19\% & 3.04\%
\end{tabular}
    \end{center}
  \end{table}

So far, we have described how to generalize the methodology from
Section~\ref{sec:2} to the case where non-zero default frequencies have been
recorded. In the following section we investigate the impact of non-zero
default correlation on the PD estimates that are effected by applying the \emph{most
  prudent estimation} methodology.

\section{Example: Correlated Default Events}
\label{sec:4}
In this section, we describe the dependence of the default events
with the one-factor probit model\footnote{According to De
Finetti's theorem \citep[see, e.g.,][Theorem (6.8)]{Durrett96},
assuming one systematic factor only is not very restrictive.} that
was the starting point for developing the
risk weight functions given in \citet{BC04a}\footnote{%
See \citet{Gordy03} and \citet{BC04b} for the background of the
risk weight functions. In the case of non-zero realized default
rates \citet{Balthazar03} uses the one-factor model for deriving
confidence intervals of the PDs. }. First, we use the example from
Section~\ref{sec:2} and assume that no default at all was observed
in the whole portfolio during the last period.
%
%
In order to illustrate the effects of correlation, we apply the
minimum value of the asset correlation that appears in the Basel
II corporate risk weight function. This minimum value is $12\%$
\citep[][paragraph 272]{BC04a}. Our model, however, works with any
other correlation assumption as well. Likewise, the \emph{most
prudent estimation} principle could potentially be applied to
others than the Basel II type credit risk model as long as the
inequalities can be solved for $p_A$, $p_B$ and $p_C$,
respectively.

Under the assumptions of this section, the confidence region at
level $\gamma$ for $p_A$ is represented as the set of all
admissible values of $p_A$ that satisfy the inequality
\citep[cf.][Sections 2.1.2 and 2.5.1 for the derivation]{Bluhm03}
\begin{equation}
  \label{eq:corrA}
  1-\gamma \ \le \ \int_{-\infty}^\infty \varphi(y)\,\left(1-\Phi\bigl(
                   \frac{\Phi^{-1}(p_A)-\sqrt{\rho}\,y}{\sqrt{1-\rho}}\bigr)
                 \right)^{n_A+n_B+n_C}\,d\,y,
\end{equation}
where $\varphi$ and $\Phi$ stand for the standard normal density
and standard normal distribution function, respectively.
$\Phi^{-1}$ denotes the inverse function of $\Phi$, and $\rho$ is
the \emph{asset correlation} (here $\rho$ is chosen as $\rho
=12\%$). Similarly to \eqref{eq:2a}, the right-hand side of
inequality \eqref{eq:corrA} tells us the one-period probability of
not observing any default among $n_A+ n_B + n_C$ obligors with
average PD $p_A$.

Solving\footnote{%
See Appendix~\ref{sec:appendix}, Proposition~\ref{pr:2} for
additional information. Taking into account correlations entails
an increase in numerical complexity. Therefore, it might seem to
be more efficient to deal with the correlation problem by choosing
an appropriately enlarged confidence level in the independent
default events approach as described in Sections~\ref{sec:2} and
\ref{sec:3}. However, it remains open how a confidence level for
the uncorrelated case, that ``appropriately'' adjusts for the
correlations, can be derived. } Equation \eqref{eq:corrA}
numerically\footnote{%
The more intricate calculations for this paper were conducted by means of the
software R \citep[cf.][]{R03}.
} for the frequencies as given in
\eqref{eq:3} leads to Table~\ref{tab:7} with maximum solutions
$\hat{p}_A$ of \eqref{eq:corrA} for different confidence levels
$\gamma$.
\begin{table}[htbp]
    \begin{center}
\parbox{\captionwidth}{\caption{ \label{tab:7}\emph{Upper confidence bound
      $\hat{p}_A$ of $p_A$, $\hat{p}_B$ of $p_B$ and $\hat{p}_C$ of $p_C$ as a
function of the
  confidence level $\gamma$. No defaults observed,
frequencies of obligors in
  grades given by \eqref{eq:3}. Case of correlated default events.}}}
\begin{tabular}{c||c|c|c|c|c|c}
$\gamma$ & 50\% & 75\% & 90\% & 95\% & 99\% & 99.9\% \\ \hline
$\hat{p}_A$ & 0.15\% & 0.40\% & 0.86\% & 1.31\% & 2.65\% & 5.29\%
\\ \hline
$\hat{p}_B$ & 0.17\% & 0.45\% & 0.96\% & 1.45\% & 2.92\% & 5.77\%
\\ \hline
$\hat{p}_C$ &  0.37\% & 0.92\% & 1.89\% & 2.78\% & 5.30\% & 9.84\%
\end{tabular}
    \end{center}
  \end{table}

Comparing the values from the first line of Table \ref{tab:7} with
Table \ref{tab:1} shows that the impact of taking care of
correlations is moderate for the low confidence levels 50\% and
75\%. The impact is much higher for the levels higher than 90\%
(for the confidence level 99.9\% the bound is even six times
larger). This observation reflects the general fact that
introducing unidirectional stochastic dependence in a sum of
random variables entails a redistribution of probability mass from
the center of the distribution towards its lower and upper limits.

The formulae for the estimations of upper confidence bounds for
$p_B$ and $p_C$ can be derived analogously to \eqref{eq:corrA} (in
combination with \eqref{eq:4a} and \eqref{eq:5a}). This yields the
inequalities
\begin{subequations}
\begin{align}
\label{eq:corrB}
  1-\gamma & \le  \int_{-\infty}^\infty \varphi(y)\,\left(1-\Phi\bigl(
                   \frac{\Phi^{-1}(p_B)-\sqrt{\rho}\,y}{\sqrt{1-\rho}}\bigr)
                 \right)^{n_B+n_C}\,d\,y\\
  \intertext{and}
\label{eq:corrC}
1-\gamma & \le  \int_{-\infty}^\infty
\varphi(y)\,\left(1-\Phi\bigl(
                   \frac{\Phi^{-1}(p_C)-\sqrt{\rho}\,y}{\sqrt{1-\rho}}\bigr)
                 \right)^{n_C}\,d\,y,
\end{align}
\end{subequations}
to be solved for $p_B$ and $p_C$ respectively. The numerical
calculations with \eqref{eq:corrB} and \eqref{eq:corrC} do not
deliver additional qualitative insights. For the sake of
completeness, however, the maximum solutions $\hat{p}_B$ of
\eqref{eq:corrB} and $\hat{p}_C$ of \eqref{eq:corrC} for different
confidence levels $\gamma$ are listed in lines 2 and 3 of Table
\ref{tab:7}, respectively.

Second, we apply our correlated model to the example from
Section~\ref{sec:3} and assume that three defaults were observed
during the last period. In analogy to Equations~\eqref{eq:p_A},
\eqref{eq:3defs} and \eqref{eq:corrA}, the confidence region at
level $\gamma$ for $p_A$ is represented as the set of all values
of $p_A$ that satisfy the inequality
\begin{subequations}
\begin{align}
  \label{eq:3defscorr}
  1-\gamma & \le  \int_{-\infty}^\infty \varphi(y)\,\sum_{i=0}^3
\left(\begin{smallmatrix}
  n_A+n_B+n_C\\ i
\end{smallmatrix}\right)
\,G(p_A,\rho, y)^i\,(1-G(p_A,\rho, y))^{n_A+n_B+n_C-i}\,d\,y,
\intertext{where the function $G$ is defined by}
\label{eq:G}
    G(p,\rho, y) & =
    \frac{\Phi^{-1}(p)-\sqrt{\rho}\,y}{\sqrt{1-\rho}}.
\end{align}
\end{subequations}
Solving \eqref{eq:3defscorr} for $\hat{p}_A$ with obligor frequencies
as given in \eqref{eq:3}, and the respective modified equations for
$\hat{p}_B$ and $\hat{p}_C$ yields the following results:

\begin{table}[htbp]
    \begin{center}
\parbox{\captionwidth}{\caption{ \label{tab:8}\emph{Upper confidence bound
      $\hat{p}_A$ of $p_A$, $\hat{p}_B$ of $p_B$ and $\hat{p}_C$ of $p_C$ as a
function of the   confidence level $\gamma$. No default observed
in grade $A$, two defaults observed in grade $B$, one default
observed in grade $C$, frequencies of obligors in grades given by
\eqref{eq:3}. Case of correlated default events.}}}
\begin{tabular}{c||c|c|c|c|c|c}
$\gamma$ & 50\% & 75\% & 90\% & 95\% & 99\% & 99.9\% \\ \hline
$\hat{p}_A$ & 0.72\% &  1.42\% &  2.50\% &  3.42\% &  5.88\% &
10.08\% \\ \hline $\hat{p}_B$ & 0.81\% &  1.59\% &  2.77\% &
3.77\% & 6.43\% & 10.92\% \\ \hline $\hat{p}_C$ & 0.84\% &  1.76\%
& 3.19\% & 4.41\% &  7.68\% & 13.14\%
\end{tabular}
    \end{center}
  \end{table}

Not surprisingly, the maximum solutions for $\hat{p}_A$,
$\hat{p}_B$ and $\hat{p}_C$ increase if we introduce defaults
in our example. Other than that, the results do not deliver
essential additional insights.

\section{Potential Extension: Calibration by Scaling Factors}
\label{sec:scaling}
One of the drawbacks of the \emph{most prudent
estimation} principle is that in the few defaults case, for all grades the
upper confidence bound PD
estimates are higher than the average default rate
of the overall portfolio. This phenomenon is not surprising, given that
we include all defaults of the overall portfolio in the upper
confidence bound estimation even for the highest rating grade.
However, these estimates might be regarded as too conservative by some
practitioners.

A potential remedy would be a scaling\footnote{%
A similar scaling procedure has recently been suggested by \citet{Cathcart05}.
} of all of our estimates
towards the central tendency (the average portfolio default rate).
We introduce a scaling factor $K$ to our estimates such that the
overall portfolio default rate is exactly met, i.e.
\begin{equation}
  \label{eq:K}
  {\textstyle\frac{\hat{p}_A \,n_A + \hat{p}_B \, n_B + \hat{p}_C \,
n_C}{n_A+n_B+n_C}} \,  K\ =\ PD_{\text{Portfolio}}.
\end{equation}
The new, scaled PD estimates will then be
\begin{equation}
  \label{eq:scaledPD}
  \hat{p}_{X, \text{scaled}} = K \, \hat{p}_X, \qquad X = A, B, C.
\end{equation}
The results of the application of such a scaling factor to our few
defaults examples of Sections \ref{sec:3} and \ref{sec:4} are
shown in Tables \ref{tab:9} and \ref{tab:10}, respectively.
\begin{table}[ht]
    \begin{center}
\parbox{\captionwidth}{\caption{ \label{tab:9}\emph{Upper confidence bounds
      $\hat{p}_{A, \text{scaled}}$ of $p_A$, $\hat{p}_{B, \text{scaled}}$ of $p_B$
      and $\hat{p}_{C, \text{scaled}}$ of $p_C$ as a
function of the confidence level $\gamma$ after scaling to the
central tendency. No default observed in grade $A$, two defaults
observed in grade $B$, one default observed in grade $C$,
frequencies of obligors in grades given by \eqref{eq:3}. Case of
uncorrelated default events.}}}
\begin{tabular}{c||c|c|c|c|c|c}
$\gamma$ & 50\% & 75\% & 90\% & 95\% & 99\% & 99.9\% \\ \hline
Central tendency
& 0.375\% &  0.375\% &  0.375\% &  0.375\% &  0.375\% & 0.375\% \\
\hline
$K$ & 0.71 &  0.48 &  0.35 & 0.30 & 0.22 & 0.17
\\ \hline
$\hat{p}_{A, \text{scaled}}$ & 0.33\% &  0.31\% &  0.29\% & 0.29\% &
0.28\% & 0.27\% \\ \hline
$\hat{p}_{B, \text{scaled}}$ &
0.37\% &  0.35\% &  0.34\% & 0.33\% & 0.32\% & 0.31\% \\
\hline $\hat{p}_{C, \text{scaled}}$ & 0.40\% &  0.43\% & 0.46\% & 0.47\%
& 0.49\% & 0.50\%
\end{tabular}
    \end{center}
  \end{table}
\begin{table}[ht]
    \begin{center}
\parbox{\captionwidth}{\caption{ \label{tab:10}\emph{Upper confidence bounds
      $\hat{p}_{A, \text{scaled}}$ of $p_A$, $\hat{p}_{B, \text{scaled}}$
      of $p_B$ and $\hat{p}_{C, \text{scaled}}$ of $p_C$ as a
function of the confidence level $\gamma$ after scaling to the
central tendency. No default observed in grade $A$, two defaults
observed in grade $B$, one default observed in grade $C$,
frequencies of obligors in grades given by \eqref{eq:3}. Case of
correlated default events.}}}
\begin{tabular}{c||c|c|c|c|c|c}
$\gamma$ & 50\% & 75\% & 90\% & 95\% & 99\% & 99.9\% \\ \hline
Central tendency
& 0.375\% &  0.375\% &  0.375\% &  0.375\% &  0.375\% & 0.375\% \\
\hline $K$ & 0.46 &  0.23 &  0.13 & 0.09 & 0.05 & 0.03
\\ \hline
$\hat{p}_{A, \text{scaled}}$ & 0.33\% &  0.33\% &  0.32\% & 0.32\% &
0.32\% & 0.32\% \\ \hline
$\hat{p}_{B, \text{scaled}}$ &
0.38\% &  0.37\% &  0.36\% & 0.36\% & 0.35\% & 0.35\% \\
\hline $\hat{p}_{C, \text{scaled}}$ & 0.39\% &  0.40\% & 0.41\% & 0.42\%
& 0.42\% & 0.42\%
\end{tabular}
    \end{center}
  \end{table}

The average estimated portfolio PD will now fit exactly the
overall portfolio central tendency. Thus, we loose all
conservatism in our estimations. Given the poor default data base
in typical applications of our methodology, this might be seen as
a disadvantage rather than an advantage. By using the \emph{most
prudent estimation} principle to derive ``relative'' PDs before
scaling them down to the final results, we preserve however the
sole dependence of the PD estimates upon the borrower frequencies in
the respective rating grades, as well as the monotony of the PDs.

There remains the question of the appropriate
confidence level for above calculation. Although the average
estimated portfolio PD now always fits the overall portfolio
default rate, the confidence level determines the ``distribution''
of that rate over the rating grades. In above example, though, the
differences in distribution appear small, especially in the
correlated case, such that we would not explore this issue
further. The confidence level could, in practice, be used to
control for the spread of PD estimates over the rating grades --
the higher the confidence level, the higher the spread.

However, above scaling only works if there is a non-zero number of
defaults in the overall portfolio. Zero default portfolios would
indeed be treated worse if we continue to apply our original
proposal to them, compared to using scaled PDs for low default
portfolios.

A variant of above scaling proposal, that takes care of both
issues, is the use of an upper confidence bound for the
overall portfolio PD in lieu of the actual default rate. This
upper confidence bound for the overall portfolio PD, incidently,
equals the \emph{most prudent estimate} for the highest rating
grade. Then, the same scaling methodology as described above can
be applied. The results of its application to the few defaults examples
as in Tables \ref{tab:9} and \ref{tab:10} are presented in Tables \ref{tab:ann2_2}
and \ref{tab:ann2_4}.
\begin{table}[htbp]
    \begin{center}
\parbox{\captionwidth}{\caption{ \label{tab:ann2_2}\emph{Upper
confidence bounds
      $\hat{p}_{A, \text{scaled}}$ of $p_A$, $\hat{p}_{B, \text{scaled}}$ of
      $p_B$ and $\hat{p}_{C, \text{scaled}}$ of $p_C$ as a
function of the confidence level $\gamma$ after scaling to the
upper confidence bound of the overall portfolio PD. No default
observed in grade $A$, two defaults observed in grade $B$, one
default observed in grade $C$, frequencies of obligors in grades
given by \eqref{eq:3}. Case of uncorrelated default events.}}}
\begin{tabular}{c||c|c|c|c|c|c}
$\gamma$ & 50\% & 75\% & 90\% & 95\% & 99\% & 99.9\% \\ \hline
Upper bound for portfolio PD
& 0.46\% &  0.65\% &  0.83\% &  0.97\% &  1.25\% & 1.62\% \\
\hline $K$ & 0.87 &  0.83 &  0.78 & 0.77 & 0.74 & 0.71
\\ \hline
$\hat{p}_{A, \text{scaled}}$ & 0.40\% &  0.54\% &  0.65\% & 0.74\% &
0.92\% & 1.16\% \\ \hline $\hat{p}_{B, \text{scaled}}$ &
0.45\% &  0.61\% &  0.74\% & 0.84\% & 1.06\% & 1.32\% \\
\hline $\hat{p}_{C, \text{scaled}}$ & 0.49\% &  0.75\% & 1.01\% & 1.22\%
& 1.62\% & 2.17\%
\end{tabular}
    \end{center}
  \end{table}
\begin{table}[htbp]
    \begin{center}
\parbox{\captionwidth}{\caption{ \label{tab:ann2_4}\emph{Upper confidence bounds
      $\hat{p}_{A, \text{scaled}}$ of $p_A$, $\hat{p}_{B, \text{scaled}}$ of $p_B$
      and $\hat{p}_{C, \text{scaled}}$ of $p_C$ as a
function of the confidence level $\gamma$ after scaling to the
upper confidence bound of the overall portfolio PD. No default
observed in grade $A$, two defaults observed in grade $B$, one
default observed in grade $C$, frequencies of obligors in grades
given by \eqref{eq:3}. Case of correlated default events.}}}
\begin{tabular}{c||c|c|c|c|c|c}
$\gamma$ & 50\% & 75\% & 90\% & 95\% & 99\% & 99.9\% \\ \hline
Upper bound for portfolio PD
& 0.71\% &  1.42\% &  2.50\% &  3.42\% &  5.88\% & 10.08\% \\
\hline $K$ & 0.89 &  0.87 &  0.86 & 0.86 & 0.86 & 0.87
\\ \hline
$\hat{p}_{A, \text{scaled}}$ & 0.64\% &  1.24\% &  2.16\% & 2.95\% &
5.06\% & 8.72\% \\ \hline $\hat{p}_{B, \text{scaled}}$ &
0.72\% &  1.38\% &  2.39\% & 3.25\% & 5.54\% & 9.54\% \\
\hline $\hat{p}_{C, \text{scaled}}$ & 0.75\% &  1.53\% & 2.76\% & 3.80\%
& 6.61\% & 11.37\%
\end{tabular}
    \end{center}
  \end{table}

As, in contrast to the situation of Tables \ref{tab:9} and \ref{tab:10},
in Tables \ref{tab:ann2_2}
and \ref{tab:ann2_4} the overall default rate in the portfolio
depends on the confidence level, we observe scaled PD estimates for the grades that increase
with growing levels. Nevertheless, the scaled PD estimates for the better grades
are still considerably lower than
the corresponding unscaled estimates from Sections \ref{sec:3} and \ref{sec:4}, respectively.
For the sake of comparison, we provide in Annex
\ref{sec:appendix2} the analogous numerical results for
the no default case.

The advantage of this latter variant of the scaling approach is that the degree of conservatism
is actively manageable by the appropriate choice of the confidence
level for the estimation of the upper confidence bound of the overall
portfolio PD. Moreover, it works for both the zero default and the
few defaults case, and thus does not produce a structural break
between both scenarios. Lastly, the results are less conservative
than the ones of our original methodology.

Consequently, we would propose to use the \emph{most prudent
estimation} principle to derive ``relative'' PDs over the rating
grades, and subsequently scale them down according to the upper
bound of the overall portfolio PD, which is once more determined by the
\emph{most prudent estimation principle} with an appropriate confidence level.

\section{Potential Extension: The multi-period case}
\label{sec:multiperiod}
So far, we have only considered the situation where estimations
are carried out on a one year (or one observation period) data
sample. In case of a time series with data from several years, the
PDs (per rating grade) for the single years could be estimated and
could then be used for calculating weighted averages of the PDs in
order to make more efficient use of the data. Proceeding this way,
however, the interpretation of the estimates as upper confidence
bounds at some pre-defined level would be lost.

Alternatively, the data of all years could be pooled and tackled
as in the one-year case. When assuming cross-sectional and
intertemporal independence of the default events, the methodology
as presented in Sections \ref{sec:2} and \ref{sec:3} can be
applied to the data pool by replacing the one-year frequency of a grade
with the sum of the frequencies of this grade over the years (analogous
for the numbers of defaulted obligors). This way, the
interpretation of the results as upper confidence bounds as well
as the frequency-dependent degree of conservatism of the estimates will
be preserved.

However, when turning to the case of default events which are
cross-sectionally and intertemporally correlated, pooling does not
allow for an adequate modelling. An example would be a portfolio
of long-term loans, where in the intertemporal pool every obligor
would appear several times. As a consequence, the dependence
structure of the pool would have to be specified very carefully,
as the structure of correlation over time and of cross-sectional
correlation are likely to differ.

In this section, we present a multi-period extension of the
cross-sectional one-factor correlation model that has been
introduced in Section \ref{sec:4}. We will take the perspective of
an observer of a cohort of obligors over a fixed interval of time.
The advantage of such a view arises from the possible conceptional
separation of time and cross-section effects.
Again, we do not present the methodology in full generality but
rather introduce it by way of an example.

As in Section \ref{sec:4}, we assume that, at the beginning of the
observation period, we have got $n_A$ obligors in grade $A$, $n_B$
obligors in grade $B$, and $n_C$ obligors in grade $C$. In
contrast to Section \ref{sec:4}, the length of the observation
period this time is $T > 1$. We consider only the obligors that
were present at the beginning of the observation period. Any
obligors entering the portfolio afterwards are neglected for the
purpose of our estimation exercise. Nevertheless, the number of
observed obligors may vary from year to year as soon as any
defaults occur.

As in the previous sections, we first consider the estimation of
the PD $p_A$ for grade $A$. PD in this section denotes a long-term
average one-year probability of default. Working again with the
\emph{most prudent estimation principle}, we assume that the PDs
$p_A$, $p_B$, and $p_C$ are equal, i.e.\ $p_A = p_B = p_C = p$. We
assume, in the spirit of \citet{Gordy03}, that a default of obligor
$i = 1, \ldots, N = n_A + n_B + n_C$ in year $t = 1, \ldots, T$ is
triggered if the change in value of their assets results in a
value lower than some default threshold $c$ as described below
(Equation \eqref{eq:prob}). Specifically, if $V_{i, t}$ denotes
the change in value of obligor $i$'s assets, $V_{i, t}$ is given
by
\begin{equation}\label{eq:V}
    V_{i, t}\ =\ \sqrt{\rho}\,S_t + \sqrt{1-\rho}\,\xi_{i, t},
\end{equation}
where $\rho$ stands for the \emph{asset correlation} as introduced
in Section \ref{sec:4}, $S_t$ is the realisation of the
\emph{systematic factor} in year $t$, and $\xi_{i, t}$ denotes the
\emph{idiosyncratic} component of the change in value. The
cross-sectional dependence of the default events stems from the
presence of the systematic factor $S_t$ in all the
obligors' change in value variables. Obligor $i$'s default occurs
in year $t$ if
\begin{equation}\label{eq:occ}
    V_{i, 1} > c,\ \ldots,\ V_{i, t-1} > c,\ V_{i, t} \le c.
\end{equation}
The probability
\begin{equation}\label{eq:prob}
    \mathrm{P}[V_{i, t} \le c]\ = \ p_{i,t}\ =  \ p
\end{equation}
is the parameter we are interested to estimate: It describes the
long-term average one-year probability of default among the
obligors that have not defaulted before. The indices $i$ and $t$
at $p_{i,t}$ can be dropped because by the assumptions we are
going to specify below $p_{i,t}$ will neither depend on $i$ nor on
$t$. To some extent, therefore, $p$ may be considered a
\emph{through-the-cycle} PD.

For the sake of computational feasibility, and in order to keep as
close as possible to the Basel~II risk weight model, we specify
the factor variables $S_t, t = 1, \ldots, T,$ and $\xi_{i, t}, i=
1, \ldots, N, t=1, \ldots, T,$ as standard normally distributed
\citep[cf.][]{Bluhm03}. Moreover, we assume that the random vector
$(S_1,\ldots, S_T)$ and the random variables $\xi_{i, t}, i= 1,
\ldots, N, t=1, \ldots, T,$ are independent. As a consequence,
from \eqref{eq:V} follows that the change in value variables
$V_{i, t}$ are all standard normally distributed. Therefore,
\eqref{eq:prob} implies that the default
threshold\footnote{%
At first sight, the fact that in our model the default threshold
is constant over time seems to imply that the model does not
reflect the possibility of rating migrations. However, by
construction of the model, the \emph{conditional} default
threshold at time $t$ given the value $V_{i, t-1}$ will in
general differ from $c$. As we make use of the joint distribution
of the $V_{i, t}$, therefore rating migrations are implicitly
taken into account.
} $c$ is determined by
\begin{equation}\label{eq:c}
    c \ =\ \Phi^{-1}(p),
\end{equation}
with $\Phi$ denoting the standard normal distribution function.

While the single components $S_t$ of the vector of systematic
factors generate the cross-sectional correlation of the default
events at time $t$, their intertemporal correlation is effected by
the dependence structure of the factors $S_1, \ldots, S_T$. We
further assume that not only the components but also the vector as
a whole is normally distributed. Since the components of the
vector are standardized, its joint distribution is completely
determined by the correlation matrix
\begin{subequations}
\begin{equation}\label{eq:intertempcorr}
     \left(%
\begin{array}{ccccc}
  1 & r_{1,2} & r_{1,3} & \cdots & r_{1,T}\\
  r_{2,1} & 1 & r_{2,3} & \cdots & r_{2,T} \\
  \vdots &  & \ddots &  & \vdots \\
  r_{T-1, 1} & \cdots & r_{T-1, T-2} & 1 & r_{T-1, T} \\
  r_{T, 1} & \cdots & r_{T ,T-2} & r_{T, T-1} & 1 \\
\end{array}%
\right).
\end{equation}
Whereas the cross-sectional correlation within one year is
constant for any pair of obligors, empirical observation indicates
that the effect of intertemporal correlation becomes weaker with
increasing distance in time. We express this distance-dependent
behavior\footnote{%
\citet{Blochwitzetal04} proposed the specification of the
intertemporal dependence structure according to
\eqref{eq:timecorr} for the purpose of default probability
estimation. } of correlations by setting in
\eqref{eq:intertempcorr}
\begin{equation}\label{eq:timecorr}
    r_{s, t} \ = \ \vartheta^{|s-t|}, \quad s, t = 1, \ldots, T, s
    \neq t,
\end{equation}
\end{subequations}
for some appropriate $0 < \vartheta < 1$ to be specified below.

Let us assume that within the $T$ years observation period $k_A$
defaults were observed among the obligors that were initially
graded $A$, $k_B$ defaults among the initially graded $B$ obligors
and $k_C$ defaults among the initially graded $C$ obligors. For
the estimation of $p_A$ according to the most prudent estimation
principle, therefore we have to take into account $k =
k_A+k_B+k_C$ defaults among $N$ obligors over $T$ years. For any
given confidence level $\gamma$, we have to determine the maximum
value $\hat{p}$ of all the parameters $p$ such that the inequality
\begin{equation}\label{eq:nodetails}
    1-\gamma\ \le \ \mathrm{P}[\text{No more than $k$ defaults observed}]
\end{equation}
is satisfied -- note that the right-hand side of
\eqref{eq:nodetails} depends on the one-period probability of
default $p$. In order to derive a formulation that is accessible
to numerical calculation, we have to rewrite the right-hand side
of \eqref{eq:nodetails}.

As the first step we develop an
expression for obligor $i$'s conditional probability to default
during the observation period, given a realization of the
systematic factors $S_1, \ldots, S_T$.
From \eqref{eq:V}, \eqref{eq:occ}, \eqref{eq:c} and by using of the
conditional independence of the $V_{i, 1}, \ldots, V_{i, T}$ given
the systematic factors, we obtain
\begin{eqnarray}
\mathrm{P}[\text{Oligor $i$ defaults}\,|\,S_1, \ldots, S_T] & = &
\mathrm{P}\bigl[\min_{t=1, \ldots, T} V_{i, t} \le
        \Phi^{-1}(p)\,|\,S_1, \ldots, S_T\bigr]\notag\\
        & = & 1 - \mathrm{P}[\xi_{i, 1} > G(p, \rho, S_1),
        \ldots,\xi_{i, T} > G(p, \rho, S_T)\,|\,S_1, \ldots,
        S_T]\notag\\
        & = & 1- \prod_{t=1}^{T}\bigl(1-\Phi(G(p,\rho,S_t))\bigr),\label{eq:oneobligor}
\end{eqnarray}
where the function $G$ is defined as in \eqref{eq:G}.
By construction, in our model all the probabilities
$\mathrm{P}[\text{Oligor $i$ defaults}\,|\,S_1, \ldots, S_T]$ are
equal, so that, for any of the $i$s, we can define
\begin{subequations}
\begin{equation}\label{eq:pi}
\begin{split}
 \pi(S_1, \ldots, S_T) & = \mathrm{P}[\text{Oligor $i$
defaults}\,|\,S_1, \ldots, S_T]\\
        & = 1- \prod_{t=1}^{T}\bigl(1-\Phi(G(p,\rho,S_t))\bigr).
\end{split}
\end{equation}
Using this abbreviation, we can write the right-hand side of
\eqref{eq:nodetails} as
\begin{equation}\label{eq:disintegration}
\begin{split}
    \mathrm{P}[\text{No more than $k$ defaults observed}] & =
    \sum_{\ell=0}^{k} \mathrm{E}\bigl[ \mathrm{P}[
    \text{Exactly $\ell$ obligors default}\,|\,S_1, \ldots,
    S_T]\bigr]\\
    & = \sum_{\ell=0}^{k} \left(%
\begin{smallmatrix}
  N \\
  \ell
\end{smallmatrix}%
\right) \mathrm{E}\bigl[ \pi(S_1, \ldots,
S_T)^{\ell}\,\bigl(1-\pi(S_1, \ldots, S_T)\bigr)^{N-\ell}\bigr]
\end{split}
\end{equation}
The expectations in \eqref{eq:disintegration} are expectations
with respect to the random vector $(S_1, \ldots,
    S_T)$ and have to be calculated as
$T$-dimensional integrals involving the density of the $T$-variate
standard normal distribution with correlation matrix given by
\eqref{eq:intertempcorr} and \eqref{eq:timecorr}. When solving \eqref{eq:nodetails} for
$\hat{p}$, we calculated the values of these $T$-dimensional integrals by means of
Monte-Carlo simulation, taking advantage of the fact that the
term
\begin{equation}\label{eq:term}
\sum_{\ell=0}^{k} \left(%
\begin{smallmatrix}
  N \\
  \ell
\end{smallmatrix}%
\right) \pi(S_1, \ldots, S_T)^{\ell}\,\bigl(1-\pi(S_1, \ldots,
S_T)\bigr)^{N-\ell}
\end{equation}
\end{subequations}
can efficiently be evaluated by making use of \eqref{eq:30}.

In order to present some numerical results for an illustration of
how the model works, we have to fix a time horizon $T$ and values
for the cross-sectional correlation $\rho$ and the intertemporal
correlation parameter $\vartheta$. We choose $T=5$ as
\citet{BC04a} requires the credit institutions to base their PD
estimates on a time series with minimum length five years. For
$\rho$, we choose $\rho = 0.12$ as in Section \ref{sec:4}, i.e.\
again a value suggested by \citet{BC04a}. Our feeling is that
default events with a five years time distance can be regarded as
being nearly independent. Statistically, this statement might be
interpreted as something like ``the correlation of $S_1$ and $S_5$
is less than $1\%$''. Setting $\vartheta = 0.3$, we obtain
$\mathrm{corr}[S_1,\,S_5]=\vartheta^{4} = 0.81\%$. Thus, the
choice $\vartheta = 0.3$ seems to be reasonable. Note that our
choices of the parameters are purely exemplary, as to some extent
choosing the values of the parameters is rather a matter of taste
or of decisions depending on the available data or the purpose of
the estimations.

Table \ref{tab:20} shows the results of the calculations for the
case where no defaults at all were observed during five years in
the whole portfolio. The results for all the three grades are
summarized in one table. For arriving at these results,
\eqref{eq:nodetails} was first evaluated with $N = n_A + n_B +
n_C$, then with $N = n_B + n_C$, and finally with $N = n_C$. In
all three cases we set $k = 0$ in \eqref{eq:disintegration} in order to express that no
defaults were observed. Not surprisingly, the calculated
confidence bounds are much lower than those presented as in Table
\ref{tab:7}, demonstrating this way the potentially dramatic
effect of exploiting longer observation periods.
\begin{table}[htbp]
    \begin{center}
\parbox{\captionwidth}{\caption{ \label{tab:20}\emph{Upper confidence
bounds
      $\hat{p}_A$ of $p_A$, $\hat{p}_B$ of $p_B$ and $\hat{p}_C$ of $p_C$ as a
function of the
  confidence level $\gamma$. No defaults during 5 years observed,
frequencies of obligors in
  grades given by \eqref{eq:3}. Case of cross-sectionally and intertemporally
  correlated default events.}}}
\begin{tabular}{c||c|c|c|c|c|c}
$\gamma$ & 50\% & 75\% & 90\% & 95\% & 99\% & 99.9\% \\ \hline
$\hat{p}_A$ & 0.03\% &    0.06\% &   0.11\% &   0.16\% &   0.30\%
& 0.55\%
\\ \hline
$\hat{p}_B$ & 0.03\% &    0.07\% &   0.13\% &   0.18\% &   0.33\%
& 0.62\%
\\ \hline
$\hat{p}_C$ &  0.07\% &   0.14\% &   0.26\% &   0.37\% &   0.67\%
& 1.23\%
\end{tabular}
    \end{center}
  \end{table}

For Table \ref{tab:21} we did essentially the same computations as
for Table \ref{tab:20}, the difference being that we assumed that
during five years $k_A = 0$ defaults were observed in grade $A$,
$k_B =2$ defaults were observed in grade $B$, and $k_C = 1$
defaults were observed in grade $C$ (as in Sections \ref{sec:3} and \ref{sec:4} during one year).
As a consequence, we had to
set $k = 3$ in \eqref{eq:disintegration} for calculating the upper confidence bounds for $p_A$
and $p_B$, as well as $k = 1$ for the upper confidence bounds of
$p_C$. Comparing here with the results presented in Table
\ref{tab:8}, we observe again a very strong effect of taking into
account a longer time series.
\begin{table}[htbp]
    \begin{center}
\parbox{\captionwidth}{\caption{ \label{tab:21}\emph{Upper confidence
bounds
      $\hat{p}_A$ of $p_A$, $\hat{p}_B$ of $p_B$ and $\hat{p}_C$ of $p_C$ as a
function of the
  confidence level $\gamma$. During 5 years no default observed in grade
  $A$, two defaults observed in grade $B$, and one default
  observed in grade $C$.
Frequencies of obligors in
  grades given by \eqref{eq:3}. Case of cross-sectionally and intertemporally
  correlated default events.}}}
\begin{tabular}{c||c|c|c|c|c|c}
$\gamma$ & 50\% & 75\% & 90\% & 95\% & 99\% & 99.9\% \\ \hline
$\hat{p}_A$ & 0.12\% &   0.21\% &   0.33\% &   0.43\% &   0.70\% &
1.17\%
\\ \hline
$\hat{p}_B$ & 0.14\% &   0.24\% &   0.38\% &   0.49\% &   0.77\% &
1.29\%
\\ \hline
$\hat{p}_C$ &  0.15\% &   0.27\% &   0.46\% &   0.61\% &  1.01\% &
1.70\%
\end{tabular}
    \end{center}
  \end{table}

\section{Potential Applications}
The \emph{most prudent estimation} methodology described in the previous
sections can be used for a range of applications, both in a bank
internal context as well as in a Basel II context. In the latter
case, it might be of specific importance for portfolios where
neither internal nor external default data are sufficient to meet
the Basel requirements. A prime example might be Specialized
Lending. In these high-volume, low-number and low-default
portfolios, internal data are often insufficient for PD
estimations per rating grade, and might indeed even be
insufficient for central tendency estimations for the entire
portfolio (across all rating grades). Moreover, mapping to
external ratings -- although explicitly allowed in the Basel
context and widely used in bank internal applications -- might be
impossible due to the low number of externally rated exposures.

The (conservative) principle of the \emph{most prudent estimation}
could potentially serve as an alternative to the Basel slotting
approach, subject to supervisory approval. In this context, the
proposed methodology might be interpreted as a specific form of
the Basel requirement of conservative estimations in case of data
scarcity.

In a wider, bank internal context, the methodology might be used
for all sorts of low default portfolios. In particular, it could
serve as a complement to other estimation methods, whether this be
mapping to external ratings, the  proposals by
\citet{SchuermannHanson04} or others. As such, we see our proposed
methodology as one additional source for PD calibrations, that
should neither invalidate nor prejudge a bank's internal choice of
calibration methodologies.

However, we  tend to believe that our proposed methodology should
only be applied to whole rating systems and portfolios. The -- at
first sight imaginable -- calibration of PDs of individual, low
default rating grades by the most prudent estimation principle
within an otherwise data rich portfolio seems infeasible because
of the unavoidable structural break between average PDs (data rich
rating grades) and upper PD bounds (low default rating grades).
Similarly, we believe that the application of the methodology for
back-testing or similar validation tools would not add much
additional information, as the (e.g. purely expert based) average
PDs per rating grade would normally be well below our proposed
quantitative upper bounds.

\section{Open Issues}
For potential applications, a number of issues would need to be
addressed. In the following, we list the ones that seem to be the
most important to us:

\begin{itemize}
\item Which confidence levels are appropriate? The proposed most
prudent estimate could serve as a conservative proxy for
\emph{average} PDs. In determining the confidence level, the
impact of a potential underestimation of these average PDs should
be taken into account. One might think that the transformation of
average PDs into some kind of ``stress'' PDs, as done in the Basel
II and many other credit risk models, could justify rather low
confidence levels for the PD estimation in the first place (i.e.
using the models as providers of additional buffers against
uncertainty). However, this conclusion would be misleading, as it
mixes two different types of ``stresses'': the Basel II model
``stress'' of the single systematic factor over time, and the
estimation uncertainty ``stress'' of the PD estimations.

Nevertheless, we would argue for moderate confidence levels when
applying the \emph{most prudent estimation} principle, but
according to another reasoning: The most common alternative to our
methodology, namely deriving PDs from averages of historical
default rates per rating grade, yields a comparable probability of
underestimating the true PD. As such, high confidence levels in
our methodology would be hard to justify.
%
%
\item At which number of defaults should one deviate from our
methodology and use ``normal'' average PD estimation methods (at
least for the overall portfolio central tendency)? Can this
critical number be analytically determined?
\item If the relative number of defaults in one of the better
ratings grades is significantly higher than those in lower rating
grades (and within low default portfolios, this might happen with
only one or two additional defaults), then our PD estimates can
turn out to be non-monotone. In which cases should this be taken
as an indication for the non-correctness of the ordinal ranking?
Certainly, monotony or non-monotony of our upper PD bounds do not
immediately imply that the average PDs are monotone or
non-monotone.
Under which conditions would there be statistical evidence of a
violation of the monotony requirement for the PDs?
\end{itemize}

Currently, we do not have definite solutions to above issues. We
believe, though, that some of them will involve a certain amount
of expert judgment rather than analytical solutions. In
particular, that might be the case with the first item. If our
proposed approach would be used in a supervisory -- say Basel II
-- context, supervisors might want to think about suitable
confidence levels that should be consistently applied.

\section{Conclusions}
In this article, we have introduced a methodology for estimating
probabilities of default in low or no default portfolios. The
methodology is based on upper confidence intervals by use of the
\emph{most prudent estimation}.
Our methodology uses all available quantitative information. In
the extreme case of no defaults in the entire portfolio, this
information consists solely of the absolute numbers of
counter-parties per rating grade.

The lack of defaults in the entire portfolio prevents
\emph{reliable} quantitative statements on both the absolute level
of \emph{average} PDs per rating grade as well as on the relative risk
increase from rating grade to rating grade. Within the \emph{most
prudent estimation} methodology, we do not use such information.
The only additional assumption used is the \emph{ordinal} ranking
of the borrowers, which is assumed to be correct.

Our PD estimates might seem rather high at first sight. However,
given the amount of information that is actually available, the
results do not appear out of range. We believe that the
choice of moderate confidence levels is appropriate within most
applications. The results can be scaled to any appropriate central tendency.
Additionally, the multi-year context as described
in Section \ref{sec:multiperiod} might provide further insight.

\textbf{Acknowledgment.} The authors thank Til Schuermann,
Claudia Sand, and two anonymous referees for providing
useful hints on an earlier draft of
this paper.


\appendix

\section{Appendix}
\label{sec:appendix}

This appendix provides some additional information on the
analytical and numerical solutions of Equations \eqref{eq:3defs}
and \eqref{eq:corrA}.

\textbf{Analytical solution of Equation \eqref{eq:3defs}.} If $X$
is a binomially distributed random variable with size
parameter $n$ and success probability $p$, then for any integer
$0\le k\le n$ we have
\begin{equation}\label{eq:30}
    \sum_{i=0}^k
\left(\begin{smallmatrix}
  n\\ i
\end{smallmatrix}\right)
\,p^i\,(1-p)^{n-i} = \mathrm{P}[X\le k] = 1-\mathrm{P}[Y\le p] =
\frac{\int_p^1 t^k\, (1-t)^{n-k-1}\, dt} {\int_0^1 t^k\,
(1-t)^{n-k-1}\, dt},
\end{equation}
with Y denoting a beta distributed random variable with parameters
$\alpha=k+1$ and $\beta= n-k$ \citep[see, e.g.,][Lemma
11.2]{Hinderer80}. The beta distribution function and its inverse
function are available in standard numerical tools, e.g.\ in
Excel.

\textbf{Direct numerical solution of Equation \eqref{eq:3defs}.}
The following proposition shows the existence and uniqueness of
the solution of \eqref{eq:3defs}, and, at the same time, provides
us with initial values for the numerical root-finding (see \eqref{eq:a.3}).
\begin{proposition} \label{pr:1}
Let $0 \le k < n$ be integers, and define the function $f_{n,k}: (0,1) \to
\mathbb{R}$ by
\begin{subequations}
\begin{equation}
  \label{eq:a.1}
  f_{n,k}(p) \ =\ \sum_{i=0}^k \left(\begin{smallmatrix}
  n\\ i
\end{smallmatrix}\right) \,p^i\,(1-p)^{n-i}, \qquad p\in(0,1).
\end{equation}
Fix some $0 < v <1$. Then the equation
\begin{equation}
  \label{eq:a.2}
  f_{n,k}(p) \ = \ v
\end{equation}
has exactly one solution $0 < p = p(v) < 1$. Moreover, this solution $p(v)$
satisfies the inequalities
\begin{equation}
  \label{eq:a.3}
1 - \sqrt[n]{v} \ \le\ p(v) \ \le \ \sqrt[n]{1-v}.
\end{equation}
\end{subequations}
\end{proposition}
\textbf{Proof.} A straight-forward calculation yields
\begin{equation}
  \label{eq:a.4}
\frac{d \,f_{n,k}(p)}{d\,p}  \ = \ - (n-k)
\left(\begin{smallmatrix}
  n\\ k
\end{smallmatrix}\right) p^k\,(1-p)^{n-k-1}.
\end{equation}
Hence $f_{n,k}$ is strictly decreasing. This implies uniqueness of the
solution of \eqref{eq:a.2}. The inequalities
\begin{equation}
  \label{eq:a.5}
  f_{n,0}(p)\ \le \ f_{n,k}(p) \ \le\ f_{n,n-1}(p)
\end{equation}
imply existence of a solution of \eqref{eq:a.2} and the
inequalities \eqref{eq:a.3}. \hfill $\Box$
\begin{table}[b]
    \begin{center}
\parbox{\captionwidth}{\caption{ \label{tab:ann2_1}\emph{Upper confidence bounds
      $\hat{p}_{A, scaled}$ of $p_A$, $\hat{p}_{B, scaled}$ of $p_B$
      and $\hat{p}_{C, scaled}$ of $p_C$ as a
function of the confidence level $\gamma$ after scaling to the
upper confidence bound of the overall portfolio PD. No default
observed, frequencies of obligors in grades given by \eqref{eq:3}.
Case of uncorrelated default events.}}}
\begin{tabular}{c||c|c|c|c|c|c}
$\gamma$ & 50\% & 75\% & 90\% & 95\% & 99\% & 99.9\% \\ \hline
Upper bound for portfolio PD
& 0.09\% &  0.17\% &  0.29\% &  0.37\% &  0.57\% & 0.86\% \\
\hline $K$ & 0.61 &  0.66 &  0.60 & 0.58 & 0.59 & 0.59
\\ \hline
$\hat{p}_{A, \text{scaled}}$ & 0.05\% &  0.11\% &  0.17\% & 0.22\% &
0.33\% & 0.51\% \\ \hline $\hat{p}_{B, \text{scaled}}$ &
0.06\% &  0.13\% &  0.20\% & 0.25\% & 0.39\% & 0.58\% \\
\hline $\hat{p}_{C, \text{scaled}}$ & 0.14\% &  0.24\% & 0.45\% & 0.58\%
& 0.89\% & 1.35\%
\end{tabular}
    \end{center}
  \end{table}

\textbf{Numerical solution of Equation \eqref{eq:corrA}.} For
\eqref{eq:corrA} we can derive a result similar to Proposition
\ref{pr:1}. However, there does not exist an obvious upper bound
to the solution $p(v)$ of \eqref{eq:a11} as in \eqref{eq:a.3}.
\begin{proposition}
  \label{pr:2}
For any probability $0 <p<1$, any correlation $0<\rho<1$ and any
real number $y$ define
\begin{subequations}
  \begin{equation}
    \label{eq:a10}
    F_\rho(p, y) \ = \ \Phi\bigl(
                   \frac{\Phi^{-1}(p)+\sqrt{\rho}\,y}{\sqrt{1-\rho}}\bigr),
  \end{equation}
where we make use of the same notations as for Equation \eqref{eq:corrA}.
Fix a value $0<v<1$ and a positive integer $n$. Then the equation
\begin{equation}
  \label{eq:a11}
  v \ = \ \int_{-\infty}^\infty \varphi(y) \bigl(1- F_\rho(p, y)\bigr)^n\, d y,
\end{equation}
with $\varphi$ denoting the standard normal density,
has exactly one solution $0 < p = p(v) < 1$. This solution $p(v)$ satisfies
the inequality
\begin{equation}
  \label{eq:a12}
  p(v) \ \ge \ 1-\sqrt[n]{v}.
\end{equation}
\end{subequations}
\end{proposition}
\textbf{Proof of Proposition \ref{pr:1}.} Note that -- for fixed $\rho$ and $y$ -- the
function $F_\rho(p, y)$ is strictly increasing and continuous in
$p$. Moreover, we have
\begin{equation}
  \label{eq:a13}
 0 \,=\,\lim_{p\to 0} F_\rho(p, y)\qquad\text{and}\qquad 1 \,=\,\lim_{p\to 1} F_\rho(p, y).
\end{equation}
Equation \eqref{eq:a13} implies existence and uniqueness of the solution of
\eqref{eq:a11}.

Define the random variable $Z$ by
\begin{equation}
  \label{eq:a144}
  Z \ = \ F_\rho(p, Y),
\end{equation}
where $Y$ denotes a standard normally distributed random variable. Then $Z$
has the well-known \emph{Vasicek distribution} \citep[cf.][]{Vasicek}, and in
particular we have
\begin{equation}
  \label{eq:a15}
  \mathrm{E}[Z] \ = \ p.
\end{equation}
Using \eqref{eq:a144}, Equation \eqref{eq:a11} can be rewritten as
\begin{equation}
  \label{eq:a16}
  v \ = \ \mathrm{E}[(1-Z)^n].
\end{equation}
Since $y \mapsto (1-y)^n$ is convex for $0 < y<1$, by \eqref{eq:a15} Jensen's inequality
implies
\begin{equation}
  \label{eq:a17}
v \ = \ \mathrm{E}[(1-Z)^n] \ \ge \ (1- p)^n.
\end{equation}
As the right-hand side of \eqref{eq:a11} is decreasing in $p$, \eqref{eq:a12}
now follows from \eqref{eq:a17}. \hfill $\Box$
\begin{table}[b]
    \begin{center}
\parbox{\captionwidth}{\caption{ \label{tab:ann2_3}\emph{Upper confidence bounds
      $\hat{p}_{A, \text{scaled}}$ of $p_A$, $\hat{p}_{B, \text{scaled}}$ of $p_B$
      and $\hat{p}_{C, \text{scaled}}$ of $p_C$ as a
function of the confidence level $\gamma$ after scaling to the
upper confidence bound of the overall portfolio PD. No default
observed, frequencies of obligors in grades given by \eqref{eq:3}.
Case of correlated default events.}}}
\begin{tabular}{c||c|c|c|c|c|c}
$\gamma$ & 50\% & 75\% & 90\% & 95\% & 99\% & 99.9\% \\ \hline
Upper bound for portfolio PD
& 0.15\% &  0.40\% &  0.86\% &  1.31\% &  2.65\% & 5.29\% \\
\hline $K$ & 0.62 &  0.65 &  0.66 & 0.68 & 0.70 & 0.73
\\ \hline
$\hat{p}_{A, \text{scaled}}$ & 0.09\% &  0.26\% &  0.57\% & 0.89\% &
1.86\% & 3.87\% \\ \hline $\hat{p}_{B, \text{scaled}}$ &
0.11\% &  0.29\% &  0.64\% & 0.98\% & 2.05\% & 4.22\% \\
\hline $\hat{p}_{C, \text{scaled}}$ & 0.23\% &  0.59\% & 1.25\% & 1.89\%
& 3.72\% & 7.19\%
\end{tabular}
    \end{center}
  \end{table}

\section{Appendix}
\label{sec:appendix2}
This appendix provides additional numerical results for the
``scaling'' extension of
the \emph{most prudent estimation} principle according to Section
\ref{sec:scaling} in the case of no default portfolios.
In the examples presented in Tables \ref{tab:ann2_1} and \ref{tab:ann2_3},
the confidence level for deriving the
upper confidence bound for the overall portfolio PD, and the
confidence levels for the \emph{most prudent estimates} of PDs
per rating grade have always been set equal. Moreover, our
methodology always provides equality between the upper bound of
the overall portfolio PD and the \emph{most prudent estimate} for
$p_A$ according to the respective examples of Sections \ref{sec:2}
and \ref{sec:4}.

\end{document}